Thermodynamic approach for enhancing superconducting critical current performance

Masashi Miura[1,2,3], Go Tsuchiya[1], Takumu Harada[1], Keita Sakuma[1], Hodaka Kurokawa[4], Naoto Sekiya[5], Yasuyuki Kato[6], Ryuji Yoshida[7], Takeharu Kato[7], Koichi Nakaoka[8], Teruo Izumi[8], Fuyuki Nabeshima[4], Atsutaka Maeda[4], Tatsumori Okada[9], Satoshi Awaji[9], Leonardo Civale[2] and Boris Maiorov[2]

Correspondence: Masashi Miura (masashi-m@st.seikei.ac.jp)

[1]*Graduate School of Science and Technology, Seikei University, 3-3-1 Kichijoji-kitamachi, Musashino-shi, Tokyo 180-8633, Japan*, [2]*Materials Physics and Applications Division, Los Alamos National Laboratory, Los Alamos, New Mexico 87545, USA*, [3]*JST-FOREST, 7, Gobancho, Chiyoda-ku, Tokyo 102-0076 Japan*, [4]*Department of Basic Science, The University of Tokyo, Meguro, Tokyo 153-8902, Japan*, [5]*Department of Electrical Engineering, University of Yamanashi, 4-3-11 Takeda, Kofu 400-8511, Japan,* [6]*Department of Applied Physics, The University of Tokyo, Bunkyo, Tokyo 113-8656, Japan,* [7]*Nanostructures Research Laboratory, Japan Fine Ceramics Center, 2-4-1 Mutsuno, Atuta-ku, Nagoya 456-8587, Japan,* [8]*National Institute of Advanced Industrial Science and Technology, 1-2-1 Namiki, Tsukuba, Ibaraki 305-8564, Japan,* [9]*Institute for Materials Research, Tohoku University, Katahira 2-1-1, Aoba-ku, Sendai 980-8577, Japan*

# Abstract

The addition of artificial pinning centers has led to an impressive increase in critical current density ($J_c$) in a superconductor, enabling record-breaking all-superconducting magnets and other applications. $J_c$ has reached~ 0.2-0.3 $J_d$, where $J_d$ is the depairing current density, and the numerical factor depends on the pinning optimization. By modifying $\lambda$ and/or $\xi$, the penetration depth and coherence length, respectively, we can increase $J_d$.

For $(Y_{0.77}Gd_{0.23})Ba_2Cu_3O_y$ ((Y,Gd)123) we achieve this by controlling the carrier density, which is related to $\lambda$ and $\xi$. We also tune $\lambda$ and $\xi$ by controlling the chemical pressure in the Fe-based superconductors, $BaFe_2(As_{1-x}P_x)_2$ films. The variation of $\lambda$ and $\xi$ leads to an intrinsic improvement of $J_c$, via $J_d$, obtaining extremely high values of $J_c$ of 130 MA/cm$^2$ and 8.0 MA/cm$^2$ at 4.2 K, consistent with an enhancement of $J_d$ of a factor of 2 for both incoherent nanoparticle-doped (Y,Gd)123 coated conductors (CCs) and $BaFe_2(As_{1-x}P_x)_2$ films, showing that this new material design is useful to achieving high critical current densities for a wide array of superconductors. The remarkably high vortex-pinning force in combination with this thermodynamic and pinning optimization route for the (Y,Gd)123 CCs reached ~ 3.17 TN/m$^3$ at 4.2 K and 18 T (**H**||$c$), the highest values ever reported in any superconductor.

# Introduction

High temperature superconductors are attractive because their high critical temperature ($T_c$) enables them to be used at high temperature and outperform standard superconductors in terms of magnetic field performance [1, 2]. However, the limiting factor is the ability of arresting the motion of Abrikosov vortices at very high critical current. The dissipative motion of vortices can be reduced or eliminated by pinning at non-superconducting defects. There are several possible approaches to enhancing the critical current density. Over the last three decades, the enormous improvements in the properties of oxide high temperature superconductors (HTS) of the REBa$_2$Cu$_3$O$_y$ family (RE123) were mostly achieved by adding and tailoring pinning centers to immobilize vortices. The number of routes to engineer the pinning landscape to increase $J_c$ is too large to describe, and continues to be fruitful [3-12]. The creep-free $J_c$, $J_{c0,\,cal}^{NPs}(T, H)$ for strong pinning by nanoparticles ($D_{np} \geq 2\xi_{ab}$) is expressed as [11]

$$J_{c0}^{NPs} \propto N_{np} \frac{\mu_0 H_c^2 \pi \xi^2 D}{4\xi} \propto N_{np} \left(\frac{1}{\lambda^2 \xi}\right)$$

(1)

where $N_{np}$ is the density of nanoparticles (NPs), $D$ is the mean size of the NPs, $\xi_{ab}$ is the coherence length, $\lambda_{ab}$ is the London penetration depth, and $H_c$ is the thermodynamic critical field (See **SI, Section 1**). How close $J_c$ can get to the upper limit for $J_c$, i.e., the depairing current density ($J_d$), by addition of pinning centers is still an open question. The $J_d$ within the Ginzburg-Landau theory [13] is

$$J_d(T) = \frac{2\sqrt{2} H_c(T)}{3\sqrt{3} \lambda_{ab}(T)} = \frac{\phi_0}{3\sqrt{3}\pi \, \mu_0 \, \lambda_{ab}(T)^2 \, \xi_{ab}(T)} \propto \left(\frac{1}{\lambda^2 \xi}\right) \quad (2)$$

where $\phi_0$ is the flux quantum.

Experimentally, the enhancement of $J_c$ by tuning the carrier density, especially in standard RE123 films, i.e. without artificial pinning centers (APCs), has been reported [14-16]. Recently, A. Stangl et al. reported that overdoped standard Y123 films grown by Pulsed Laser Deposition (PLD) reached 18% of $J_d$ at 5 K, which is a consequence of the increase of the condensation energy with charge carrier density [17]. On the other hand, by adding and tailoring of APCs, the highest $J_c$ achieved in RE123 and Fe-based films is in the range of 10-20% of $J_d$ [4-6, 8-13]. Most of the studies introducing APCs in RE123 are of the coherent $BaMO_3$(BMO, M=Zr, Hf, Sn, etc.) nanorods [4, 6, 7, 9, 13] and coherent $Y_2BaCuO_5$ precipitates [5, 10]. The $c$ axis of the RE123 matrix is expanded by coherent APCs [7, 11] resulting in degradation of the carrier density due to strain-induced oxygen-vacancy formation and and loss of crystallinity [6, 7, 11]. On the other hand, we have succeeded in introducing incoherent BMO NPs into not only $(Y_{0.77}Gd_{0.23})Ba_2Cu_3O_y$ ((Y,Gd)123) films [11] but also Fe-based pnictide $BaFe_2(As_{1-x}P_x)_2$ (Ba122:P) films [8], which leaves the matrix unaltered with just a small degradation of the superconducting properties. The $BaHfO_3$ (BHO) NPs in RE123 films and $BrZrO_3$ (BZO) NPs in Ba122:P films, which are an average size ($D_{ave.}$) of 7 nm with a density $N_{np} \sim 80 \times 10^{21}$ m$^{-3}$ and $D_{ave.}$ of 8 nm with $N_{np} \sim 68 \times 10^{21}$ m$^{-3}$, respectively. For both nanocomposite materials, we have shown the large enhancement in $J_c$ at not only self-field but also in-field by introducing a high density of incoherent NPs of tailored size [8,11]. Theoretically, using the time-dependent Ginzburg-Landau equations (TDGL) and a *targeted evolution* approach, Sadovskyy *et al*. [17] explored the optimization of $J_c$, showing that a level 30-40% of $J_d$ could be attained.

Now, in addition to focusing on improving the pinning morphology, we can increase $J_d$. Considering formulas (1) and (2), we see that $J_c \propto \left(\frac{1}{\lambda^2 \xi}\right) \propto J_d$. Therefore, reducing $\xi$ or $\lambda$ would improve $J_d$ and consequently $J_c$. However, these parameters are material-specific and have not been thoroughly studied for improving $J_c$ in APC doped cuprates and Fe-based superconducting films. If both these characteristic lengths can be changed, in addition to the enhancement of flux pinning, $J_c$ can be dramatically improved through the enhancement of $J_d$. Increasing $T_c$ has been the empirical way to increase $J_d$, however; this depends on discovering new superconductors, and even in the cases where this has been achieved (e.g., $HgBa_2Ca_2Cu_3O_{8+\delta}$ and $Bi_2Sr_2Ca_2Cu_3O_{10+\delta}$) it has not led to improved performance as the gains have been negated by the enhancement in thermal fluctuations that grow as

$$G_i^{1/2} \propto (T_c^2 \gamma^2 \lambda^4 / \xi^2)^{1/2} \qquad (3)$$

[3,18] where $G_i$ is the Ginzburg number.

Here, we present a novel route to improve superconductors' performance by increasing $J_d$. Unlike the increase in pinning, which is extrinsic, this route is thermodynamic: $J_d$ is raised by decreasing $\lambda$ and/or increasing $H_c \propto (\lambda \xi)^{-1}$ [14]. This method is general and applicable to any superconductor; here we show results for RE123 and Ba122:P films both with and without incoherent BMO NPs. The method works in conjunction with any pinning landscape improvement already achieved, opening a way to increase performance independently of the microstructure. As a concomitant advantage, the decrease in $\lambda$ also reduces the deleterious effects of thermal fluctuations by reducing the $G_i$. In the RE123 compounds, we achieve this by increasing the carrier concentration and thus decreasing $\lambda$; we also detect an increase in $H_c$ and decrease of $\gamma$ observed through the increased $H_{c2}$ (*i.e.*, decrease of $\xi$) with a consequent reduction of $\gamma$. When we combine this new strategy with our previously developed methods to incorporate large density $N_{np}$ of incoherent BHO NPs of tailored size, we obtain $J_c \sim 150$ MA/cm$^2$ ($\sim 32.4\%$ of $J_d$) and $J_c \sim 130$ MA/cm$^2$ ($\sim 28\%$ of $J_d$) at 4.2 K and self-field for nanocomposite (Y,Gd)123 films on single-crystal substrates and metallic substrates (coated conductors), respectively. These improvements carry over to the *in-field* properties. We also apply this route in Ba122:P films with incoherent BZO NPs where we can increase $J_d$ and $J_c$ by controlling $\lambda$ and $\gamma$ through tuning of the chemical pressure. This coordinated strategy can inform the improvement efforts in the newly discovered hydrogen-based superconductors.

We start by increasing $J_d$ by decreasing $\lambda$ and $\xi$ for RE123. The Cu-O planes containing chains in RE123 are an exception among other cuprates (La$_{2-x}$Sr$_x$CuO$_4$ [19], Y$_{1-x}$Cu$_x$Sr$_2$Cu$_2$Tl$_{0.5}$Pb$_{0.5}$O$_7$ [20], Tl$_2$Ba$_2$CuO$_{6+x}$ [21, 22]) or Fe-pnictides [18]. This gives a unique ability for $\lambda$ to be decreased [23] and $H_c$ to be increased [24] up to the highest possible overdoping, unlike other cuprates [21,25] and Fe-pnictides [26] for which $\lambda$ is minimized at the optimum $T_c$ doping. An indication of the possible gain in terms of enhancing $J_d$ by tuning the carrier concentration is observed in the specific-heat jump (directly related to $H_c$) that for $y=7$ ($p=0.19$) is 45% higher than for optimum doping. These beneficial effects outweigh the negative effects of the 4% decrease of $T_c$. Thus, we proceed to change the oxygen content $y$ and modify the carrier concentration $p$, to ultimately change $\lambda$ and $\xi$ for RE123 with two very different pinning landscapes. In **Table 1**, we summarize the main experimental results for RE123 compounds. To avoid changes stemming from different $T_c$ values, we compare two samples with similar $T_c$ (89.2 and 90.2K) that are on either sides of the optimal doping, i.e., $p=0.18$ and 0.144. The comparison of these samples leads to $J_d$ of 498 and 230 MA/cm$^2$ respectively, almost a factor of two increase with the decrease of $\lambda$ and $\xi$.

## Materials and Methods

### The film growth

The epitaxially grown Y123 nanocomposite films of standard (Y,Gd)123 and BHO NP doped (Y,Gd)123 ((Y,Gd)123+BHO) films were grown from metal organic solutions including Y-, Gd-, and Ba-trifluoroacetates and Cu-naphthenate with the cation ratio of 0.77 : 0.23 : 1.5 : 3 on buffered tapes of CeO$_2$ (grain-boundary angles, $\Delta\phi_{CeO2}$=3.0°)/Y$_2$O$_3$/LaMnO$_3$/ion-beam-assisted deposition (IBAD)-MgO/Gd$_2$Zr$_2$O$_7$/Hastelloy C276 (Haynes International Inc., Kokomo, IN, USA). We added Hf-naphthenate into the (Y,Gd)123 solutions; the volume percent of BHO was 12, and the concentration of starting solution was 0.45 mol/L with coating thickness ($d_{coat}$) of 30 nm, result in small size (7 nm) and high density (7.5×10$^{22}$ m$^{-3}$) nanoparticles while maintaining the crystallinity and $T_c$ of the RE123 matrix. Moreover, the nanocomposite ((Y,Gd)123+BHO) films were also fabricated on CeO$_2$($\Delta\phi_{CeO2}$(220)=1.0°)/R-Al$_2$O$_3$ single crystals. For comparison, we also fabricated 10 mol.% BHO doped Eu123 (Eu123+BHO) films on buffered tapes of CeO$_2$ ($\Delta\phi_{CeO2}$(220)=3.0°)/Y$_2$O$_3$/LaMnO$_3$/ion-

beam-assisted deposition (IBAD)-MgO/Gd$_2$Zr$_2$O$_7$/Hastelloy C276. The Eu123+BHO film has coherent BHO nanorods of 5 nm diameter, not incoherent BHO NPs, which are obtained with MOD. The details of PLD film preparation have been published elsewhere [27]. The total thickness of the RE123 layer for all samples was 600 nm to 1000 nm, which was confirmed by cross sectional transmission electron microscopy (TEM (JEM-F200 (JEOL Ltd., Tokyo, Japan)). The standard BaFe$_2$(As$_{1-x}$P$_x$)$_2$ and 3 mol. % of BaZrO$_3$ (BZO) doped epitaxial films were deposited on MgO (100) single-crystal substrates by ablating the polycrystalline pulsed laser deposition targets using the second harmonic (wavelength: 532 nm) of a pulsed Nd:YAG laser at a repetition rate of 10 Hz in a vacuum of $10^{-4}$ Pa at a substrate temperature of 850°C. In this work, the amount of P substitution $x$ in the target was adjusted to be 0.33, 0.40, and 0.50. The total thickness of the (Ba122:P) films with and without BZO was 80 nm.

**The oxygenation treatments for cuprate films**

The oxygenation treatments were precisely controlled to tune the carrier concentration for the (Y,Gd)123+BHO coated conductors (CCs). The oxygenation process is reversible as confirmed by the ability to recover $T_{c,zero}$ and $J_c^{s.f.}$ after varying the oxygen content (**SI, Fig. S1** and **Table S1**). In an investigation of bulk RE123 (RE=Nd, Sm, Eu, Gd Dy, Ho and Y), the optimum doping annealing temperature ($T_A^{opt.}$) was found to depend on the RE element [28]. For maximum $T_c$, RE123 with larger RE$^{3+}$ ions needed lower O$_2$ annealing temperatures as compared to those with smaller RE$^{3+}$ ions (at the same O$_2$ pressure). The $T_A^{opt.}$ (the temperature for the highest $T_{c,zero}$ and smallest $\Delta T$ (=$T_{c,onset}$-$T_{c,zero}$) for each RE123 material) for our RE123 CCs prepared by different growth methods (MOD and PLD) is consistent with that of bulk studies [28], at $T_A^{opt.}$=500, 450 and 350°C for Y123, (Y,Gd)123 and Eu123 CCs, respectively (See **SI, Fig. S2**). In this work, for oxygenation of the RE123 films prepared from different fabrication processes (MOD and PLD), we annealed at an O$_2$ environment of 1.1 atm, and each annealing temperature ($T_A$=300-550°C) was held for 3 h, then rapidly quenched to room temperature. From the *c*-axis length measured by XRD(RINT2100 and ATX-G (Rigaku Co., Tokyo, Japan)) and the $T_{c,zero}$, annealing at 300°C, 3 h was sufficient to oxygenate the RE123 films.

**Transport properties in magnetic fields**

The films were patterned using a pulsed fiber laser (1095 nm, 20 W) into bridges of ~50 μm width. The crystalline quality was examined by x-ray diffraction (XRD). The temperature dependence of the resistivity ($\rho$) was measured by a four-probe method in the temperature range of 4-300 K using a Physical Property Measurement System (PPMS, Quantum Design, Inc., San Diego, CA, USA) with a superconducting magnet generating a field **H** up to 14 T and in an 18 T-superconducting magnet at Tohoku University. In the PPMS, a rotational stage was used to rotate the samples with respect to **H**. The critical current was determined using a 1 μV cm$^{-1}$ criterion. The $H_{c2}$ and $H_{irr}$ were determined using 0.90 $\rho_N$ and 0.01 $\rho_N$ criteria, respectively, where $\rho_N$ is the normal-state resistivity. Hall measurements were conducted in a magnetic field of 9 T. The six electrical contacts used silver paste on silver pads deposited on the film by sputtering. The magnetization studies were performed using a SQUID (Quantum Design, Inc., San Diego, CA, USA) magnetometer to characterize the temperature and field dependence of $J_c$ and *S*.

# Results

**Controlling the carrier density of the superconducting films**

First, in order to investigate the influence of introducing coherent BHO nanorods on $T_{c,zero}$, *c*-axis length and self-field $J_c$ ($J_c^{s.f.}$), we measured the hole concentration ($n_H$, determined from the Hall effect at 300 K) dependence of these properties for both standard Eu123 and Eu123 with coherent BHO nanorods (Eu123+coherent BHO) CCs grown by PLD (See **Figure 1 (a) and (b)**). $T_{c,zero}$ is determined using an 0.01$\rho_N$ criterion. As shown in the **inset of Fig.1(a)**, the Eu123+coherent BHO

CC has coherent BHO nanorods of 5 nm diameter. As shown in **Fig. 1(a)**, $T_{c,zero}$ and the *c*-axis length for standard Eu123 CCs decreases systematically with decreasing oxygenation temperature ($T_A$) from $n_H^{300K}$=9.4×10$^{21}$/cm$^3$(optimum doped) to 15.3×10$^{21}$/cm$^3$, confirming that the samples are in the overdoped regime. On the other hand, although the Eu123+coherent BHO CCs are treated at the same O$_2$ annealing condition as the overdoped standard ones, $T_{c,zero}$ is not reached even for the optimum doping level (i.e., underdoped regime) and the *c*-axis length is longer than that in standard Eu123 CCs. **Fig. 1(b)** shows $J_c^{s.f.}$ as a function of carrier concentration (doping level) in the CuO$_2$ layer (*p*) for both standard Eu123 and Eu123+coherent BHO CCs. The variation of *p* is determined by following $T_{c,zero}$ on the universal doping curve [29], where $T_{c,zero}$ reaches its maximum at optimum doping (*p*=0.16) (see **SI, Fig. S3**).   As a result, in spite of the fact that Eu123+coherent BHO films have strong pinning, $J_c^{s.f.}$ of the nanocomposite CCs is not enhanced compared to that of the standard ones because the carrier concentration *p* is lower due to the strain-induced oxygen-vacancy formation leading to degradation of the carrier doping level [7]. There are no reports of overdoped RE123 with coherent APCs because generally it is difficult to dope to the overdoped regime for coherent APC doped RE123 films, although the overdoped range is easily achieved in standard films (with no APCs).

Now we focus on our incoherent BHO NP doped (Y,Gd)123 CCs.   As shown in **Fig. 1 (c)**, in spite of the fact that (Y,Gd)123+BHO CCs have a high density of NPs (see **inset of Fig.1(c)**), $T_{c,zero}$ and the *c*-axis length systematically decrease with decreasing oxygenation temperature ($T_A$) from $n_H^{300K}$=9.4×10$^{21}$/cm$^3$(optimum doped) to 21×10$^{21}$/cm$^3$, reaching the overdoped regime. On the other hand, $J_c^{s.f.}$ at 77 K (See **Fig. 1 (d)**) increases monotonically with increasing *p* beyond optimum doping. Even though the (Y,Gd)123 and (Y,Gd)123+BHO CCs have almost the same $T_{c,zero}$-$n_H$ broad peak (because the superconducting matrix remains intact [11]), the $J_c^{s.f.}$ in the (Y,Gd)123+BHO CC is over two times higher than that in the (Y,Gd)123 CC. It is worth noting that, even though $T_{c,zero}$ ~ 90 K is almost the same for the (Y,Gd)123+BHO CCs with $n_H^{300K}$ = 7×10$^{21}$/cm$^3$ and $n_H^{300K}$ =21×10$^{21}$/cm$^3$, as shown in **Fig.1(c)**, the overdoped CC shows $J_c^{s.f.}$ 1.8 times higher at 77 K. It is worth noting that we achieved an overdoped doping level up to $p=0.18$ for our incoherent BHO NP-doped (Y,Gd)123 CCs, but not for the standard one (without APCs).

### The influence of grain boundaries in carrier density-controlled films

In order to clarify that the enhancement of $J_c$ (**Fig. 1 (d)**) for our overdoped (Y,Gd)123+BHO CCs, which are fabricated on oxide-buffered metallic substrates (in-plane crystallinity ($\Delta\phi_{CeO2}=3°$)), is not due to doping-induced improved grain boundary properties (similar to what is seen for the improvement of the intergrain-$J_c$ for Ca-doped Y123 films on bi-crystal substrates [30]), we investigated the in-plane crystallinity of $CeO_2$ ($\Delta\phi_{CeO2}$) -buffered metallic substrate-dependence of self-field $J_c$ at 77 K for underdoped (Y,Gd)123, underdoped (Y,Gd)123+$BaHfO_3$ and overdoped (Y,Gd)123+$BaHfO_3$ films (See **Figure 2 (a)**). For $\Delta\phi_{CeO2} > 3°$, $J_c^{s.f.}$ for all films decreases exponentially with increasing boundary angle. On the other hand, for $\Delta\phi_{CeO2} < 3°$, $J_c^{s.f.}$ for all films decreases by just 10% when changing $\Delta\phi_{CeO2}$ from 0.6° to 3° independent of $p$ and of pinning landscape. As shown in **Fig. 2 (b)**, although the ratio of $J_c^{s.f.}(\Delta\phi)/J_c^{s.f.}(0.6°)$ for $\Delta\phi_{CeO2} > 3°$ of two underdoped films is lower than that in an overdoped film, the ratio for $\Delta\phi_{CeO2} < 3°$ clearly shows exactly the same trend even for different $p$ (underdoped and overdoped) and for different microstructures (without and with BHO NPs). To investigate the GB misorientation angles in our MOD CCs, we study the plan-view TEM images of (Y,Gd)123 CCs on $\Delta\phi_{CeO2}=3°$ of buffered metallic substrates (See **inset of Fig. 2 (b)**). The chain of edge dislocation distances ($D$) is 13.5–15.3 nm for a film on $\Delta\phi_{CeO2}=3°$, which represents misorientation angles of 1.5–1.7° calculated using $D=(|b|/2)/\sin(\theta_{GB}/2)$, where $|b|$ is the norm of the corresponding Burgers vector. Because the crystal growth of MOD films is meandering and overgrows the substrate grain boundaries (differing from that in PLD), the value of the misorientation angle is almost half of $\Delta\phi_{CeO2}$, which is the same as the in-plane crystallinity of (Y,Gd)123 ($\Delta\phi_{(Y,Gd)123}(103)=1.5°$) as evaluated by XRD. It is clear that GBs in (Y,Gd)123 and (Y,Gd)123+$BaHfO_3$ CCs grown on $\Delta\phi_{CeO2}=3°$ of $CeO_2$ buffered metallic substrate are not Josephson weak linked (i.e., with locally suppressed order parameter). Based on **Figs. 2 (a)** and **(b)**, we conclude that our $J_c^{s.f.}$ properties for films on $\Delta\phi_{CeO2}=3°$ of $CeO_2$ buffered metallic substrates mainly depend on the value of the intragrain $J_c$, not the intergrain-$J_c$ as is the case for large misorientation angles of the GBs [30].

**The penetration depth and coherence length in carrier-controlled films**

In **Fig. 3**, we show that $\lambda$ and $\xi$ vary with carrier concentration ($p$). We observe changes consistent with those found in the literature for single-crystal samples. The **upper panel of Fig. 3(a)** presents the measured $\lambda_{ab}$ as a function of $p$ for Y123 [23, 31] and (Y,Ca)123 [32] single crystals. For refs. 23 and 31, we calculated the penetration depth using $\lambda_{ab}=[\lambda_a\lambda_b]^{1/2}$. For the extraction of $\lambda_{ab}(0)$ for our (Y,Gd)123+BHO film, we use the temperature dependence of the resonant frequency of the coplanar waveguide resonators based on the equations derived by K. Watanabe *et al.* [33] (measurement details are shown in **SI, Fig. S4** and **Fig. S5**). Indeed, for $\lambda_{ab}(0)$, the decrease with increasing $p$ for (Y,Gd)123+BHO films is similar to that of Y123 and (Y,Ca)123 single crystals[23, 31, 32].

The $H_{c2}(0)$ for **H**||*c* for the (Y,Gd)123+BHO CCs with different $p$ are shown in the **lower panel of Fig. 3(a)**. The values for (Y,Gd)123+BHO CCs and Gd123 single crystal are estimated by using the Werthamer-Helfand-Hohenberg (WHH) formula [34] (detailed data in **SI, Fig. S6** and **Fig. S7**). The (Y,Gd)123+BHO CCs with $p$=0.144, 0.168 and 0.180 exhibit an upward trend in $H_{c2}(0)$ similar to that of Gd123 single crystals (measured) and Y123 single crystals from ref [35]. The $\mu_0 H_{c2}^c(0)$ vs. $p$ trend for our (Y,Gd)123+BHO CCs and the Gd123 single crystal is consistent with Y123 single-crystal studies, showing that $H_{c2}$ increases with $p$ up to $p$=0.18.

Using $\lambda_{ab}$ (**upper panel of Fig. 3(a)**) and $\xi_{ab}$ we calculate the $H_c(0)$ using $\mu_0 H_c(0)= \frac{\phi_0}{2\sqrt{2}\pi\lambda_{ab}(0)\xi_{ab}(0)}$, for our (Y,Gd)123+BHO films and Y123 on single crystals [23, 31, 35] with various $p$, shown in the **upper panel of Fig. 3(b)**. These results are consistent with the calculated $H_c(0)$ values based on $H_{c2}$ and $H_{c1}$ for Gd123 single crystal (detailed data in **SI, Fig. S7**). $H_c(0)$ values for the (Y,Gd)123+BHO CCs increase with increasing $p$, which is a similar trend to that for the Gd123 and Y123 single crystals.

In view of the changes in $\xi$ and $\lambda$, summarized in **Table 1**, we can use $J_d \propto \left(\frac{1}{\lambda^2 \xi}\right)$ and assert that the variation of $\lambda$ is more important for $J_d$ than that of $\xi$, as $\lambda$ also affects $H_c$. The enhanced $H_c$ is one of the main reasons for the higher $J_c$ for our most overdoped (Y,Gd)123+BHO film with $p$=0.18 and $J_c$ calculated for strong pinning nanoparticles doped films, $J_{c0,cal}^{NPs} \propto \left(\frac{1}{\lambda^2 \xi}\right)$ (see **SI, Section 9**). The monotonic increase in $H_c(0)$ with $p$ shown in the **upper panel of Fig. 3(b)**, consistent with previous studies in single- [36] and poly-crystalline [24] Y123, confirms the different behavior of Y123 as compared to other HTS cuprate materials where $H_c(0)$ and $\Delta C/T_c$ coincide with the maximum of $T_c$ [20, 22].

To investigate the $n_s$ dependence of the normal-state carrier density for Y123, we calculated $n_s$ for (Y,Gd)123+BHO CCs with different $p$. The dependence of the effective mass ($m^*$) in Y123 single crystals with the hole doping level was recently measured [37]. To calculate the $n_s$ for our (Y,Gd)123+BHO CCs, we used the relationship $\lambda_{ab}(0) = \sqrt{\frac{m_*}{\mu_0 n_s(0) e^2}}$ and $m^*$ from reference [37] (no $m^*$ data above $p$=0.152 was reported, so we extrapolated from the available curve). The $n_s$ for our (Y,Gd)123+BHO CCs increases monotonically with the hole concentration, as shown in the **lower panel of Fig. 3(b)**. This is consistent with Y123 single-crystal data, also shown for comparison [23,31].

From **Figure 3**, the overdoped (Y,Gd)123+BHO CCs have larger $H_c$ and smaller $\lambda_{ab}$ as compared to the under and optimally doped CCs, indicating an increase with $p$ for $J_c$ ($\propto \lambda^{-2}\xi^{-1}$) and $J_d$ ($\propto \lambda^{-2}\xi^{-1}$). From formula (1), the (Y,Gd)123+BHO CCs with $p$=0.144, 0.168 and 0.180 yield $J_d(0)$=230, 368 and 498 MA/cm$^2$, respectively. This remarkable enhancement in $J_d(0)$ in the overdoped sample is due to the reduced $\xi_{ab}(0)$ and $\lambda_{ab}(0)$ achieved by controlling the carrier density.

**The dramatically higher $J_c$ at all temperatures**

Consistent with the enhancements of $J_c$ being induced by the changes in $J_d$, we observe the same enhancement at all temperatures over a wide range of applied magnetic fields. The **upper panel of Fig. 4(a)** shows the calculated $J_d(T)$ with different $p$ based on the parameters in **Table 1**. The overdoped (Y,Gd)123+BHO CC ($p$=0.18) has higher $J_d(T)$ as compared to films with $p$=0.168 and 0.144. As seen in the **lower panel of Fig. 4(a)**, $J_d(T)$ for $p$=0.18 is about 2 times higher than that for $p$=0.144 over a wide temperature range. **Fig. 4(b)** shows $J_c^{s.f.}(T)$ for the (Y,Gd)123+BHO CCs with different $p$. Even though all the CCs have the same high density of BHO nanoparticles, $J_c^{s.f.}$ for $p$=0.18 is the highest at all temperatures. Although CCs with $p$=0.18 and $p$=0.144 have almost the same $T_c$, slightly lower than that for the optimum doped one, $J_c^{s.f.}$ in the $p$=0.18 film is almost twice that of $p$=0.144. At $p$=0.18, the $J_c^{s.f.}$ for the (Y,Gd)123+BHO CC achieves a maximum value of 130 MA/cm$^2$ at 4.2 K. This is higher than previously reported for any superconducting material [6,9,11,13,17,38] except nanowires and ultra-thin films (where the flux pinning mechanism is different) [39-41]. The $J_c^{s.f.}$ results are independent of the sample's width ($W$) and length ($L$) as seen by the results at 77 K (**SI, Fig. S8**), indicating very uniform superconducting films. If we apply our finding to a film on a CeO$_2$ ($\Delta\phi_{ceo2}$=1°) buffered R-Al$_2$O$_3$ single crystal (shown in the inset of **Fig. 4(b)**), we achieve a $J_c^{s.f.}$ ~ 150 MA/cm$^2$ at 4.2 K, which is also the highest value ever reported for any superconducting material with thickness >> $\lambda_{ab}$. The enhancement of 1.15 in $J_c^{s.f.}$ at 4.2 K for a film on a single-crystal substrate as compared to the film on an oxide-buffered metallic substrate (all other parameters being the same) is almost the same enhancement of 1.1 at 77 K (comparing $J_c^{s.f.}$ of films on $\Delta\phi_{ceo2}$=1° and $\Delta\phi_{ceo2}$=1° of substrates in Fig.2 (b)), due to the slightly higher in-plane-crystallinity for the former.

Insights into the benefit of further controlling $J_d$ by changing the hole doping level are found in the in-field $J_c(T)$ at $\mu_0H$=0.3T and **H**∥$c$ in **Fig. 4(c)**. The overdoped (Y,Gd)123+BHO CC ($p$=0.18) shows the highest in-field $J_c$, about twice the value for $p$=0.145. As shown in the **inset of Fig. 4(c)**, the transport $J_c$ coincides very well with the $J_c$ calculated from magnetization (measured on a different piece) using the Bean model [42], indicating that our $J_c$ values are highly uniform and reproducible. The solid lines in the **upper panel of Fig. 4(c)** are the calculated $J_c$ ($J_{c0,\,cal}^{NPs}$) [3] and the solid symbols are the experimentally obtained parameters (calculation details are shown in **SI, Table S2**), indicating that $J_{c0,\,cal}^{NPs}$ for (Y,Gd)123+BHO CCs with different $p$ are in good agreement with the experimental $J_c$. This agreement confirms the critical role of the decrease in $\lambda$ and increase in $H_c$. The most interesting and important feature of the data presented here is that the enhancement ratio in $J_d$-$T$ and $J_c$-$T$ in both self and in-field are identical (see **lower panels of Fig. 4**). This confirms we can enhance $J_c$ through enhancing $J_d$ by changing $H_c$ and $\lambda$, while keeping the pinning enhancement intact.

**The Ginzburg number in carrier-controlled films**

The top panel of **Figure 5 (a)** shows $\gamma$ for (Y,Gd)123 and (Y, Gd)123+BHO CCs with various hole concentrations measured at 300 K. The $\gamma$ is calculated from the angular dependence of $H_{c2}$ (see **SI, Fig. S9** and ref. 43 for details). Both for CCs with and without BHO, the $c$-axis length (see **Fig. 1(c)**) and mass anisotropy decrease with increasing $n_H$, i.e., hole doping level. These dependences follow the same trend observed in the $c$-axis length vs. $p$ characteristics for Y123 SC [44] and in the $\gamma$ vs. $p$ characteristics for (Y,Ca)123 SC [45]. We also observe a systematic reduction of the $c$-axis length and smaller mass anisotropy for the samples with nanoparticles. The origin of this effect is under investigation and will be the focus of future publications.

As indicated above, the reduction of $\gamma$ also diminishes the effect of thermal fluctuations, as characterized by $G_i \sim \gamma^2$ (see formula (3)). While $J_c$ increases by increasing $J_d$, it can also increase by reducing the effect of flux creep. The effect of flux creep is characterized by the creep rate, $S$, with which pinned vortices escape from the pinning centers under thermal agitation. It was found that there is a universal lower limit of the creep rate $S_{min} \sim G_i^{1/2}(T/T_c)$ [18], which demonstrates that the creep rate can be essentially reduced by reducing the anisotropy of the superconductor. In addition, $S$ can also be reduced to its limit $S_{min}$ by adding pinning.

The bottom panel of **Fig. 5 (a)** shows $S(T=50K, \mu_0H=0.3T)$ vs. $n_H$ for our CCs, as a representative example of $S(T,H)$ over a wide range of conditions outside the Anderson-Kim (A-K) regime. This relationship can also be replotted as that of $S(T=T_c/4, \mu_0H=1T)$ (i.e., inside the A-K regime) vs. $G_i^{1/2}$ as shown in the **Fig. 5 (b)**. Thus reducing the anisotropy is a new way to reduce $S$. In addition, the introduction of nanoparticles is effective, as can be seen from the comparison between (Y,Gd)123+BHO CCs and (Y,Gd)123 CCs.

## Discussion

The effect of $J_d$ on $J_c$ is general and apparent for different superconductors with varied pinning landscapes (see **Fig. 6**). Here we display $J_c^{s.f.}$ as a function of $J_d$ at 4.2 K [43-52]. Details of the calculation parameters for $J_d$ and experimentally obtained $J_c$ are shown in **SI, Table S3**. First, we see that there is a general trend for several superconductors in which $J_c$ is proportional to $J_d$ (clearly seen in the **inset of Fig. 6**). Second, from **Fig. 6** for standard (Y,Gd)123 CCs, we can tune $J_c^{s.f.}$ at 4.2 K from 23.5 to 44.5 MA/cm$^2$ by changing the $J_d$. We note that the relation of $J_c \sim 0.1 J_d$ remains unchanged. The enhancements of $J_d$ and $J_c$ by tuning the carrier concentration are also observed in Ca-doped Y123 SC [50]. Third, the combination of controlling $J_d$ by tuning the carrier concentration in a film with a high density of nanoparticles leads to the highest $J_c^{s.f.}(4.2K)=130$ MA/cm$^2$ for (Y,Gd)123+BHO CCs, which is 28.0% of $J_d$. Moreover, a (Y,Gd)123+BHO film on a single crystal achieved 32.4% of $J_d$ ($J_c^{s.f.}(4.2K)=150$ MA/cm$^2$) because of the further improvement obtained from higher in-plane crystallinity. This value is close to the 33% achievement predicted by Gurevich [53] for a superconductor with NP pinning centers similar to the actual conditions in our (Y,Gd)123+BHO films and CCs.

As shown in the **inset of Fig. 6**, increasing $J_d$ is an effective way to enhance $J_c$ not only for Y123 cuprate CCs but also for Fe-based pnictide Ba122:P films. In the Ba122:P system, the isovalent substitution of P for As induces chemical pressure, suppressing magnetism and inducing superconductivity, which is different from the effect of electron doping or hole doping [54-56]. For standard Ba122:P films, by tuning the $x$, $J_c^{s.f.}$ at 4.2 K increases from 1.0 to 3.6 MA/cm$^2$ due to the increase in $J_d$ from 23 to 74 MA/cm$^2$. Moreover, Ba122:P+BZO films with different $x$ show that $J_c$ increases (up to 8 MA/cm$^2$ at 4.2 K) with increasing $J_d$, indicating that the combined approach of tuning $J_d$ and enhancing the flux pinning (i.e., adding NPs) can also be used to improve the performance of superconductors of different families. In this regard it is important to note that, although the method to control $J_d$ ($H_c$ and $\lambda_{ab}$) is different for the cuprate and for the pnictide (changing carrier concentration ($p$) and the chemical pressure ($x$), respectively), the end result is the same, thus highlighting the general applicability of our strategy.

Further insight into the effects of the combination of increasing $J_d$ and enhancing flux pinning can be obtained from the field dependence of $J_c$. **Figure 7(a)** shows $J_c(\mathbf{H}\|c)$ at 4.2 K for our overdoped (Y,Gd)123+BHO CC compared with several RE123 films and CCs [16, 57-59]. As seen for up to 18 T, the $J_c(\mathbf{H}\|c)$ of overdoped (Y,Gd)123+BHO CC is the highest among all superconductors. Compared with overdoped standard Y123 at 5 K [16], the enhancement of our overdoped (Y,Gd)123+BHO CC is 144% at self-field and 199% at 5 T. Moreover, compared to that in coherent BHO doped CCs [59], the $J_c(\mathbf{H}\|c)$ of our CC shows a 254% increase at 1 T and 175% increase at 18 T. The remarkable in-field performance of overdoped (Y,Gd)123+BHO CC is highlighted in **Fig. 7(b)**, where the pinning force, $F_p = J_c(H) \times \mu_0 H$ is compared with that of several RE123 materials [16, 57-59]. The $F_p$ at 4.2 K of our overdoped (Y,Gd)123+BHO CC reaches ~ 3.17 TN/m$^3$ at 18 T ($\mathbf{H}\|c$), which is the highest reported value for any superconductor material. Please note that because the anisotropy of RE123 materials is around 5 (see **Fig. 5(a)**), $F_p$ measured along the $c$ axis is the minimum value for a RE123 superconductor.

## Conclusions

In summary, we have succeeded in combining this thermodynamic route ($J_d$ is raised by decreasing $\lambda$ and/or increasing $H_c$) with our previously developed methods to tailor the size and incorporate large densities of incoherent nanoparticles. We obtain $J_c \sim 150$ MA/cm$^2$ (~32.4% of $J_d$) and $J_c \sim 130$ MA/cm$^2$ (~28.0% of $J_d$) at 4.2 K and self-field for nanocomposite RE123 films on single-crystal substrates and metallic substrates (CCs), respectively. Moreover, for films of chemical-pressure controlled Ba122:P with incoherent BZO NPs, $J_c^{s.f.}$ at 4.2 K increases from 1.0 to 8.0 MA/cm$^2$ due to the increase in $J_d$ in combination with the introduction of high densities of incoherent BZO NPs. To our knowledge, the $J_c$ values attained for our CC of overdoped (Y,Gd)123 at not only self-field and but also in high field are the highest reported to date for any superconductor. This highlights that thermodynamic improvements in superconductors can work in parallel with already successful artificial pinning centers, and that a maximum $J_c \sim 0.3\ J_d$ appears to be the upper limit reachable for enhancement of $J_c$, at least at this point.

## Acknowledgements

This work at Seikei University was supported by JST-FOREST (Grant Number JPMJFR202G, Japan). A part of this work at Seikei University was supported by JSPS KAKENHI (18KK0414 and 20H02184) and Promotion, Mutual Aid Corporation for Private Schools of Japan (Science Research Promotion Fund). This work at Los Alamos National Laboratory was supported by the LDRD office 20210320ER (BM) and the US DOE, Office of Basic Energy Sciences, Materials Sciences and Engineering Division (L.C.). This work at AIST was supported by the New Energy and Industrial Technology Development Organization (NEDO). We thank Jeffery O Willis (LANL) for helpful discussions and critical reading of the manuscripts. MM also thank Akira Ibi for sample fabrication of PLD-EuBCO films.

**Author Contributions**

M.M. and B.M. carried out experiment design. M.M., G.T. and T. H. grew the films and carried out transport measurement. B.M. carried out data analysis, provided advice and consultation on flux pinning and manuscript preparation. R.Y. and T.K. carried out microstructural studies. Y. K. carried out analysis for $H_{c2}$. K N. and T.I. contributed to discussion on the film preparation. H. K., F. N. and A.M. contributed to discussion on analysis for penetration depth. K.S. and N.S. carried out analysis for resonance frequency. T.O. and S.A. contributed to discussion on analysis for transport measurement at high field. L.C. contributed to discussion and manuscript preparation. All authors discussed the results and implications and commented on the manuscript. M.M. and B.M. wrote the manuscript with contribution from all the authors.

**Competing interest**

The authors declare no competing interest.

# Figure captions

**Figure 1 | Carrier concentration dependence $p$ of critical temperature $T_c$, $c$-axis length and self-field critical current density $J_c^{s.f.}$.** (a) Dependence of $T_c$ (top panel) and $c$-axis length (bottom panel) on $n_H$ at 300 K and (b) $p$ dependence of $J_c^{s.f.}$ at 77 K in standard Eu123 and Eu123+BHO CCs grown by PLD. (c) dependence of $T_c$ and $c$-axis length on $n_H$ at 300 K (f) $p$ dependence of $J_c^{s.f.}$ in standard (Y,Gd)123 and (Y,Gd)123+BHO CCs grown by MOD. Inset of Fig. 1(a) and Fig. 1(c) are cross-sectional TEM images for Eu123+BHO and (Y,Gd)123+BHO CCs, where BHO NPs are colored. Error bars in $J_c$ were determined from the uncertainty in the in-plane crystallinity, natural defects and film thickness.

**Figure 2 | Self-field critical current density $J_c^{s.f.}$ as a function of in-plane crystallinity of buffer layer ($\Delta\phi_{CeO2}=3°$).** (a) $J_c^{s.f.}$ and (b) the ratio $J_c^{s.f.}/J_c^{s.f.}(\Delta\phi_{ceo2}=0°)$ at 77 K for underdoped (Y,Gd)123, underdoped (Y,Gd)123+BaHfO$_3$ and overdoped (Y,Gd)123+BaHfO$_3$ films grown on CeO$_2$-buffered metallic substrates ($2.5° < \Delta\phi_{ceo2} < 8.5°$) and on CeO$_2$ buffered R-Al$_2$O$_3$ substrates ($0.6° < \Delta\phi_{ceo2} < 2°$). Inset of **Fig. 2(b)** shows the plan-view TEM image of (a Y,Gd)123 film on a CeO$_2$ buffered metallic substrate with $\Delta\phi_{CeO2}=3.0°$. White and yellow allows indicate the twin boundaries (TBs) and dislocations, respectively. Error bars in $J_c$ were determined from the uncertainty in the in-plane crystallinity, natural defects and film thickness.

**Figure 3 | Carrier concentration dependence $p$ of penetration depth $\lambda_{ab}$, upper critical filed $H_{c2}(0)$, thermodynamic critical field $H_c(0)$ and superfluid density $n_s(0)$.** (a) $\lambda_{ab}$ (upper panel) and $H_{c2}(0)$ estimated by using WHH (lower panel) as a function of carrier concentration ($p$) for (Y,Gd)123+BHO films. (b) Calculated $H_c(0)$ from formula (3) (upper panel) and estimated superfluid density (lower panel) as a function of $p$ for our (Y,Gd)123+BHO CCs. For reference, values for Y123[23,31,35] and (Y,Ca)123 [32] single crystals are also shown. Error bars on the $H_{c2}(0)$ for Y123 single crystal [35] are reported values which is represent the uncertainty in extrapolating the $H_{c2}(T)$ to $T=0$.

**Figure 4 | Temperature dependence of critical current density $J_c$.** (a) Calculated $J_d(T)$ for different $p$ for (Y,Gd)123+BHO CCs based on the parameters in Table 1. (b) $J_c^{s.f.}(T)$ for the (Y,Gd)123+BHO CCs with different $p$. Inset of Fig. 4(b) shows $J_c^{s.f.}(T)$ for the (Y,Gd)123+BHO ($p=0.18$) film on CeO$_2$ buffered R-Al$_2$O$_3$ single-crystal substrate. (c) In-field $J_c(T)$ at $\mu_0H=0.3$T and $\mathbf{H}\|c$ for (Y,Gd)123+BHO CCs with various $p$. Comparison of transport $J_c$ and $J_c$ calculated from

magnetization using the Bean model for (Y,Gd)123+BHO CCs is shown in the inset of Fig. 4(c).

**Figure 5 | Creep rate in overdoped (Y,Gd)123+BHO CCs.** **(a)** (top panel) calculated $\gamma$ for (Y,Gd)123 and (Y,Gd)123+BHO CCs for various hole concentrations measured at 300 K and (bottom panel) $S(T=50\text{ K}, \mu_0 H=0.3\text{ T})$ as a function of $n_H$ for (Y,Gd)123 and (Y,Gd)123+BHO CCs. **(b)** $S(T=T_c/4, \mu_0 H=1\text{T})$ vs. $G_i^{1/2}$.

**Figure 6 | Self-field critical current density $J_c^{s.f.}$ v.s. depairing current density $J_d$ for various superconducting materials.** $J_c^{s.f.}$ at 4.2 K as a function of $J_d$ at 4.2 K for different superconductors with varied pinning landscapes [43-52]. Open and solid symbol show $J_c$ for pristine and superconductors with introduced pinning centers. Inset of **Fig. 6** shows $J_c$ vs. $J_d$ for chemical-pressure controlled Fe-based pnictide Ba122:P films with and without nanoparticles.

**Figure 7 | Critical current density $J_c$ and pinning force $F_p$ as a function of magnetic field.** **(a)** $J_c(\mathbf{H}\|c)$ at 4.2 K and **(b)** $F_p$-$\mu_0 H$ curve of overdoped (Y,Gd)123+BHO CCs at 4.2 K and $\mathbf{H}\|c$. For comparison, the data for Sm123+coherent BHO film [57], (Y,Gd)123+coherent BZO CC [58], (Y,Gd)123+coherent BHO CC [59], overdoped Y123 film at 5 K [16], and Y123 CCs [58] are included.

**Table 1 | Structural and superconducting properties.** $\mu_0 H_{c2}(0)$, $\xi_{ab}(0)$, $\lambda_{ab}(0)$ and $J_d$ for (Y,Gd)123 and (Y,Gd)123+BHO CCs with different $p$. Error bars on $\mu_0 H_{c2}(0)$ were determined from the fit uncertainty in the corresponding data analysis.

| materials | $p$ | $T_c$ [K] | $\gamma = H_{c2}^{ab}/H_{c2}^{c}$ | $\mu_0 H_{c2}^{cal}(0)$ [T] | $\xi_{ab}(0)$ [nm] | $\lambda_{ab}(0)$ [nm] | $\mu_0 H_c^{cal}(0)$ [T] | $J_d(0)$ [MA/cm$^2$] |
|---|---|---|---|---|---|---|---|---|
| (Y,Gd)123 CCs | 0.177 ($T_A$=300 °C) | 90.2 | 4.7 | 119.1±1.8 | 1.66 | 114 | 1.23 | 466.61 |
| | 0.160 ($T_A$=450 °C) | 92.2 | 5.4 | 86.4±1.4 | 1.95 | 129 | 0.92 | 310.37 |
| | 0.142 ($T_A$=500 °C) | 89.8 | 5.8 | 58.4±0.9 | 2.37 | 145 | 0.68 | 201.97 |
| (Y,Gd)123+BHO CCs | 0.18 ($T_A$=300 °C) | 89.2 | 4.3 | 126.5±1.4 | 1.61 | 112 | 1.29 | 498.02 |
| | 0.168 ($T_A$=450 °C) | 91.7 | 4.7 | 97.0±0.8 | 1.84 | 122 | 1.04 | 367.68 |
| | 0.144 ($T_A$=500 °C) | 90.2 | 5.5 | 72.0±1.6 | 2.14 | 143 | 0.76 | 230.57 |

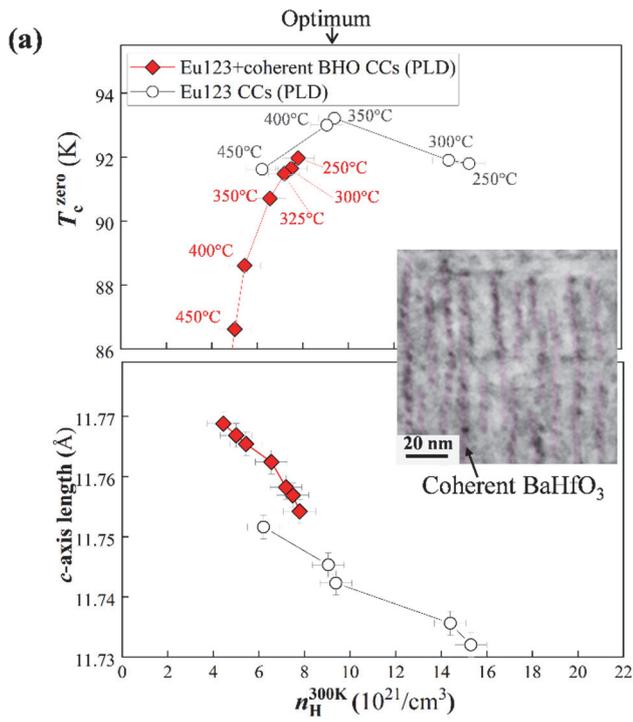
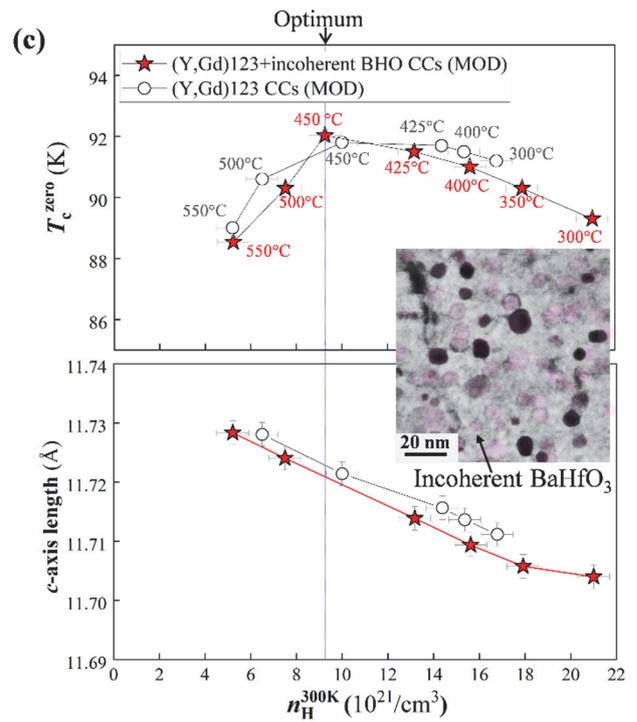
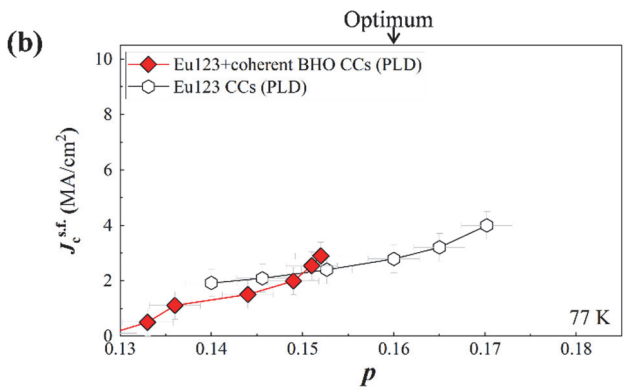
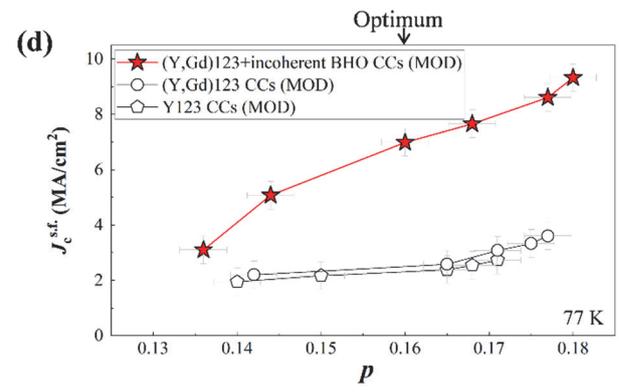

**Figure 1**

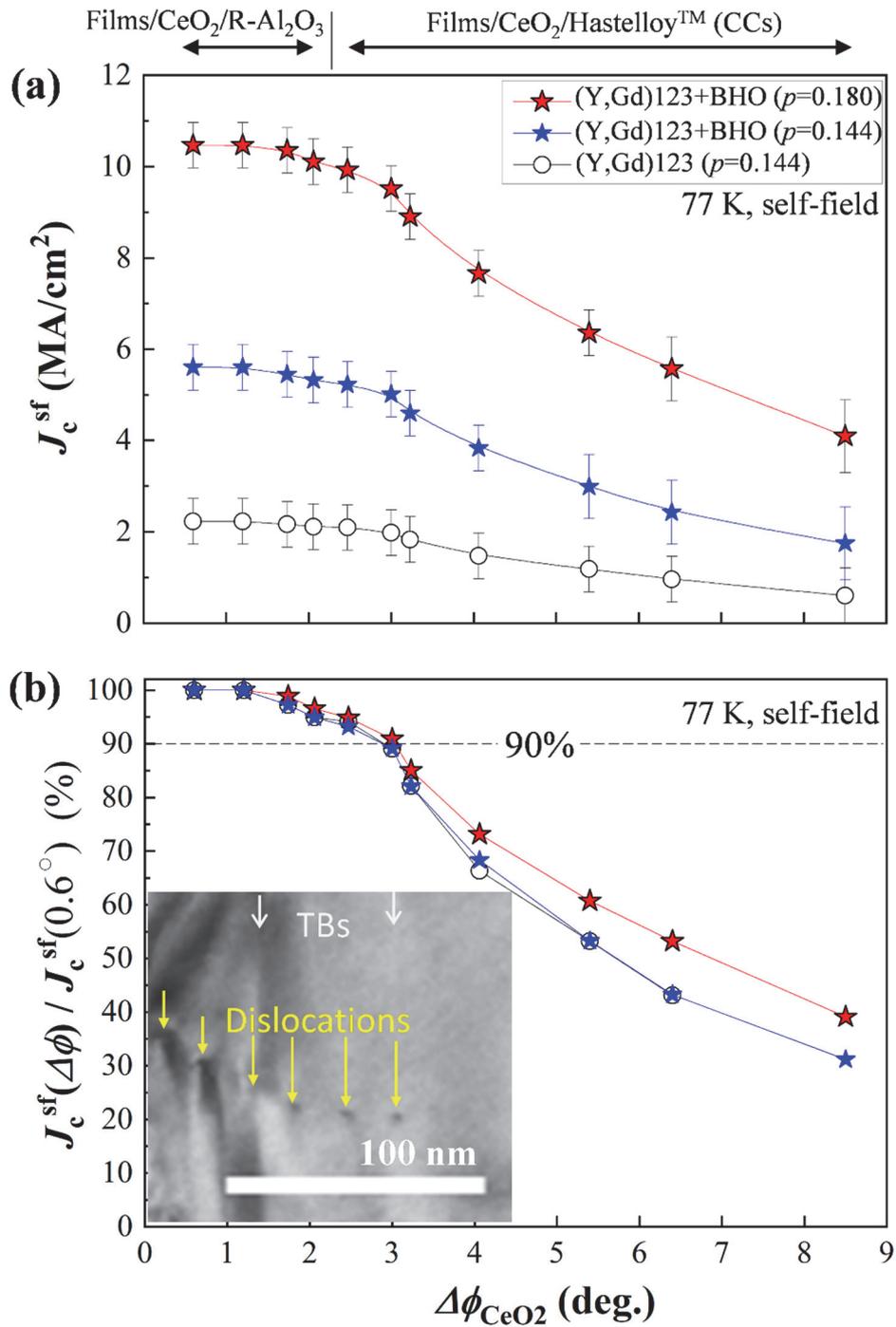

**Figure 2**

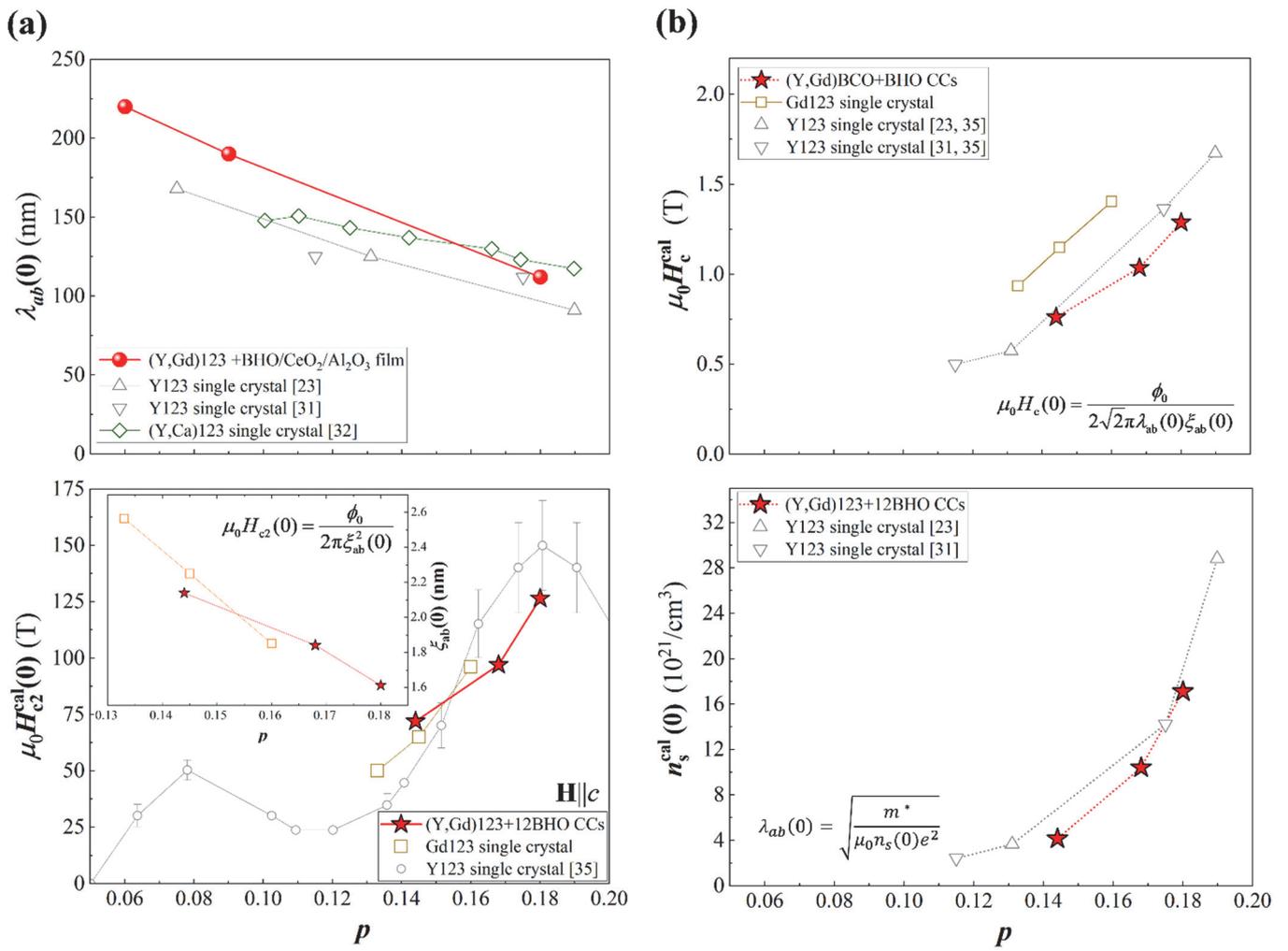

Figure 3

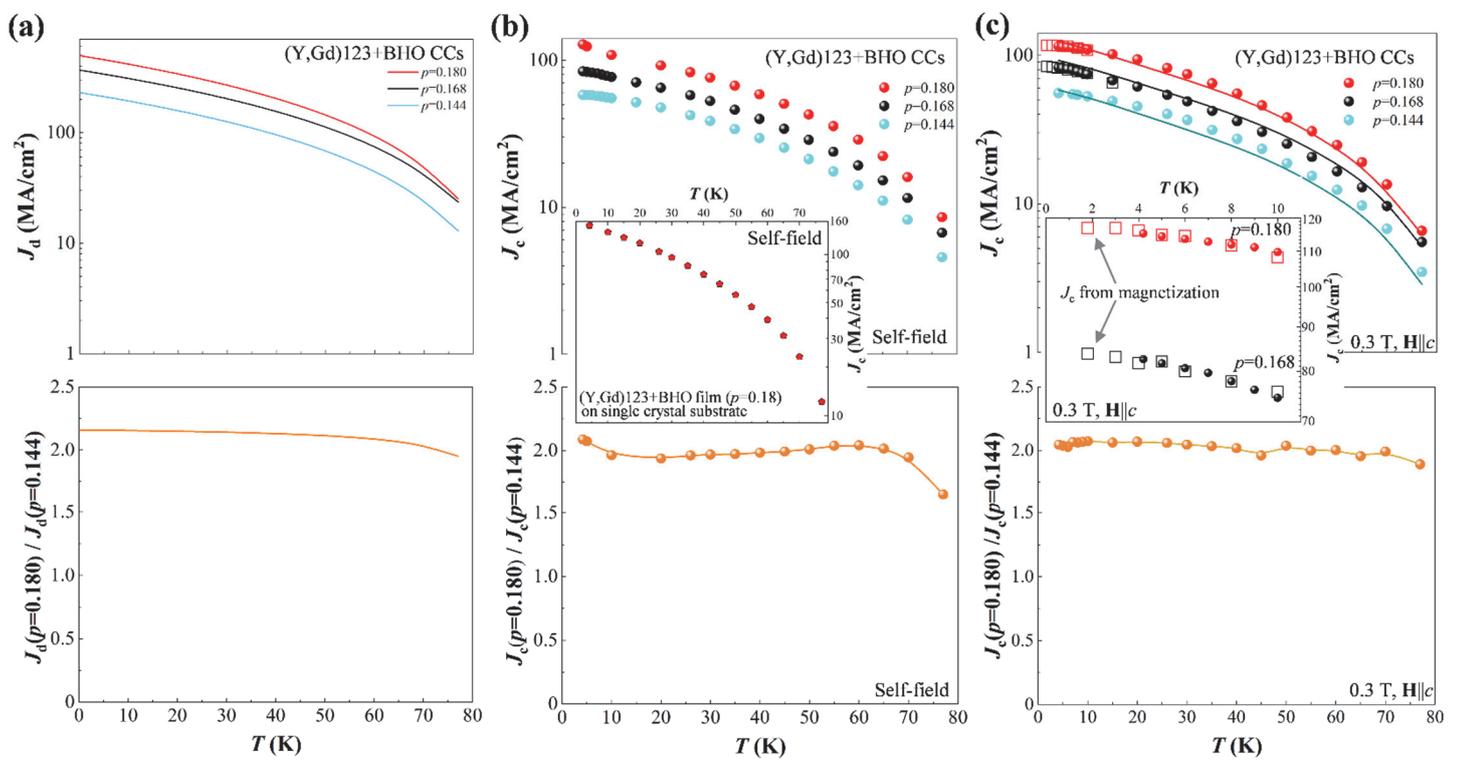

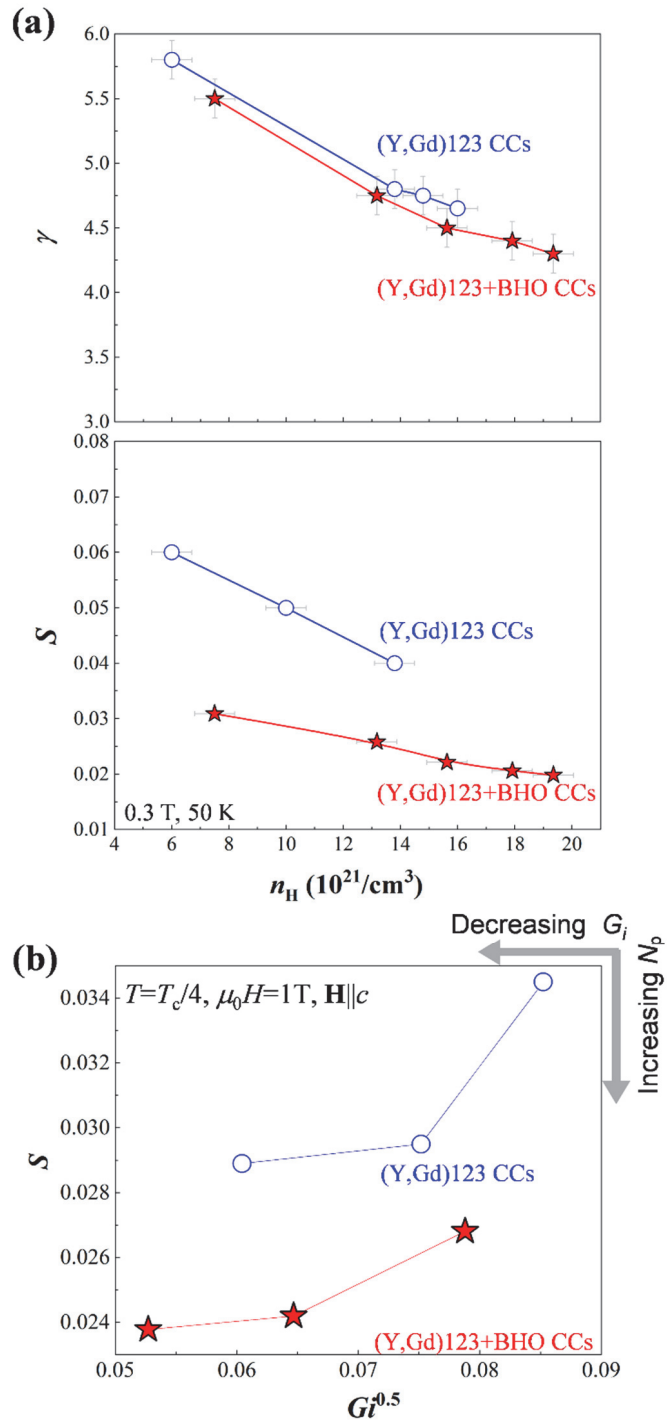

Figure 5

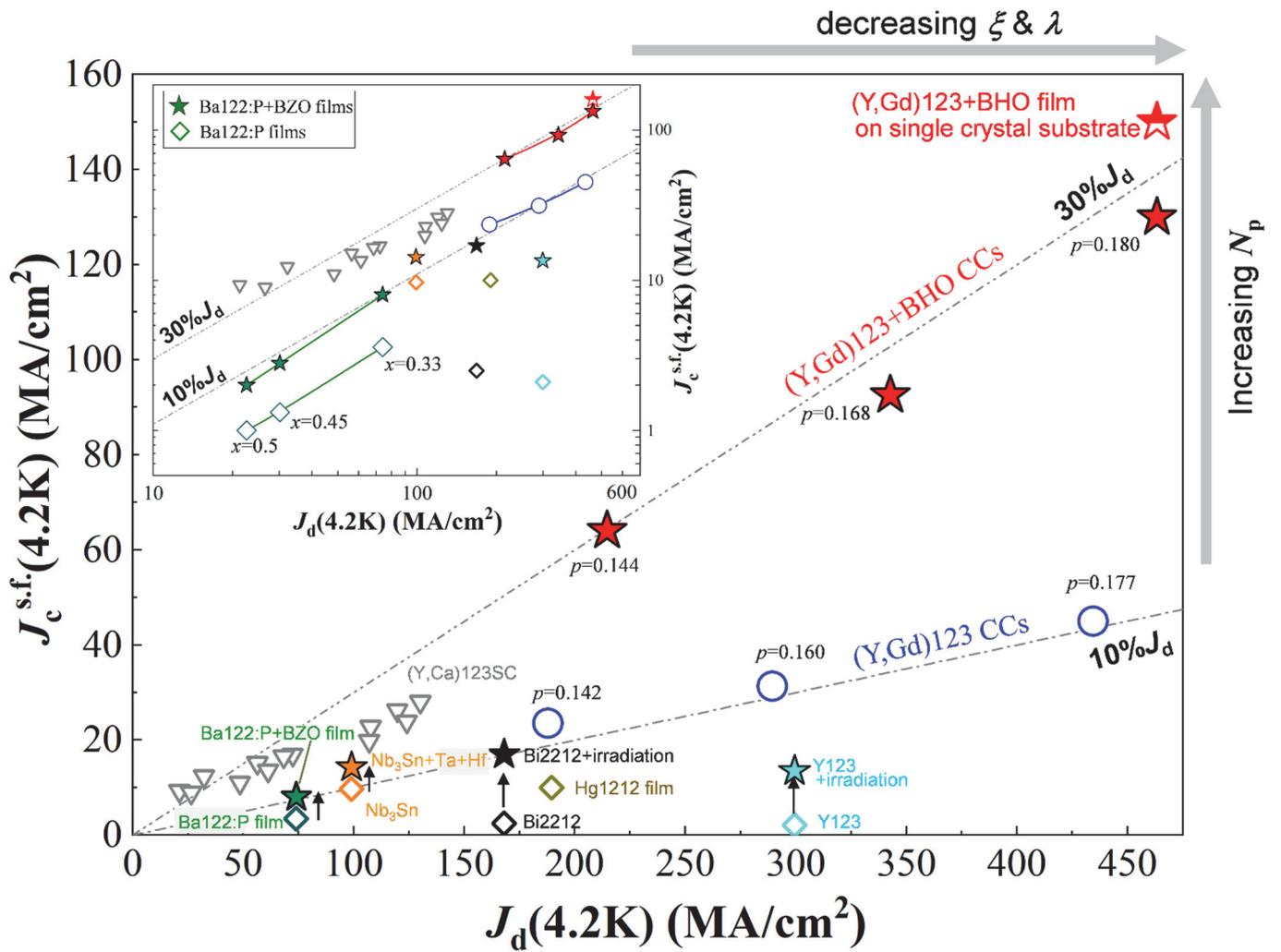

**Figure 6**

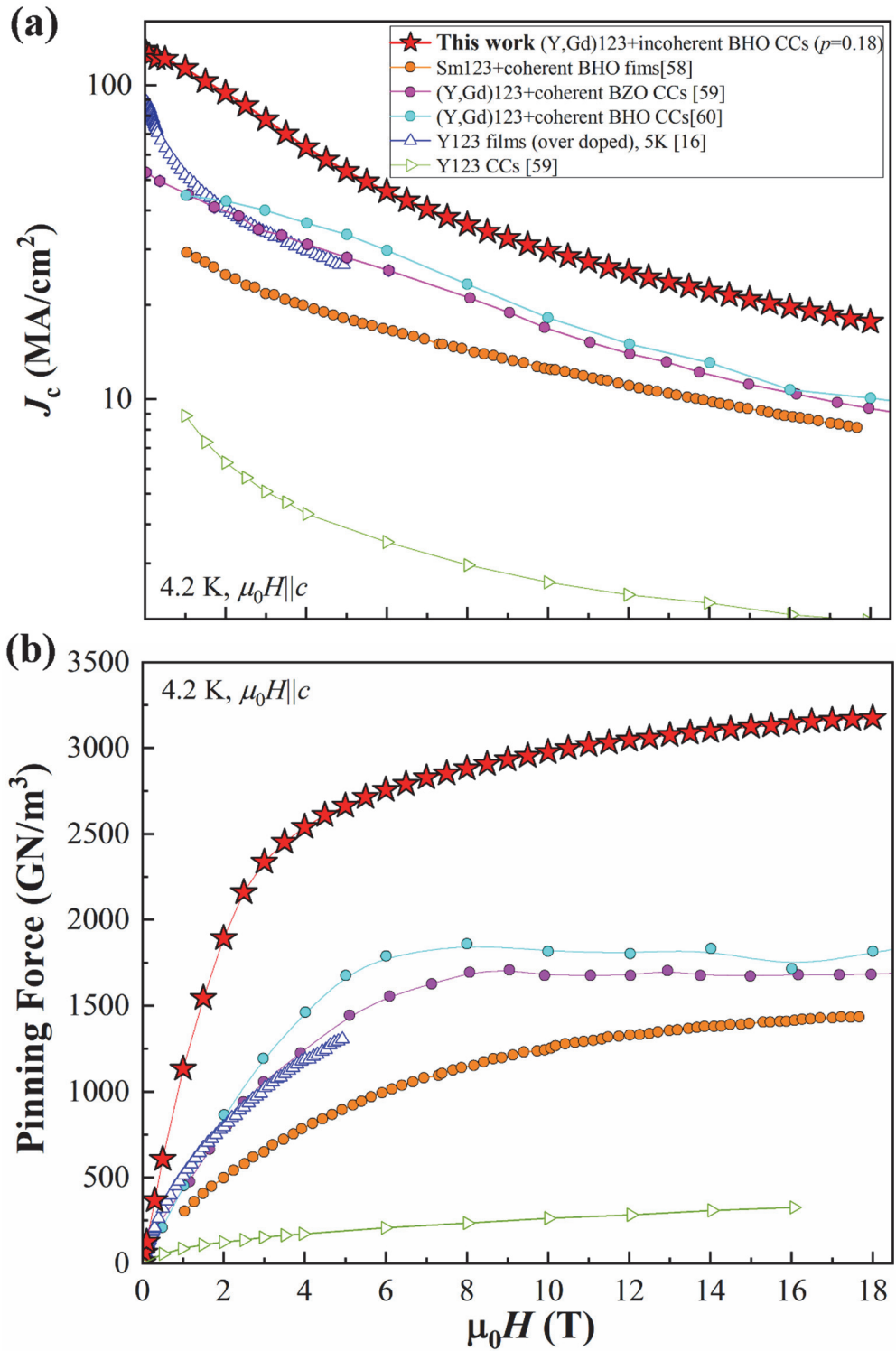

**Figure 7**